\documentclass[aps,preprint,showpacs,showkeys,nofootinbib,floatfix]{revtex4}
\usepackage{epsfig}
\usepackage{graphicx}
\begin{document} 

\title{\textbf{Light $\Xi$ hypernuclei }}
\author{H.~Garcilazo} 
\email{humberto@esfm.ipn.mx} 
\affiliation{Escuela Superior de F\' \i sica y Matem\'aticas, \\ 
Instituto Polit\'ecnico Nacional, Edificio 9, 
07738 M\'exico D.F., Mexico} 

\author{A.~Valcarce} 
\email{valcarce@usal.es} 
\affiliation{Departamento de F\'\i sica Fundamental and IUFFyM,\\ 
Universidad de Salamanca, E-37008 Salamanca, Spain}

\date{\today} 

\begin{abstract}
Arguments in favor of a light $\Xi NN$ hypernucleus with $(I)J^P=(3/2)1/2^+$
are presented, within the uncertainties of our
knowledge of the baryon-baryon strangeness $-2$ interactions. 
If bound, this $\Xi NN$ state, being decoupled from the
lowest $N\Lambda\Lambda$ system, would be stable.
It will also benefit from additional binding due to the electromagnetic 
interaction what makes it worthwhile to look for. 
We show how the equivalent state with $J=3/2$ could 
never be bound in spite of the attractive interaction of 
the two-body subsystems. We illustrate our discussion with a 
full-fledged Faddeev calculation of the $\Xi NN$ 
system using simple potentials that mimic more elaborate interactions. 
We also make contact with different recent phenomenological interactions 
from the literature, like the ESC08 Nijmegen potential or quark-model based potentials.
\end{abstract}

\maketitle 

The physics of hypernuclei is progressing dramatically in
both theory and experiment. Theoretically there have been recent proposals of the 
stability of $^4_{\Lambda\Lambda}$n~\cite{Ric15},
the existence of $\Xi$ hypernuclei~\cite{Yam10,Hiy08,Yam01}, or the strangeness $-2$
hypertriton~\cite{Gar13}. Obviously, all these predictions are subject
to the uncertainties of our knowledge of the baryon-baryon
interaction, in particular in the strangeness $-2$ sector.
Experimentally, it has been recently reported an emulsion event
providing clear evidence of a deeply bound state of the $\Xi^- -^{14}$N system~\cite{Nak15}. 
A thorough discussion of the present status of the experimental progress 
in hypernuclear physics can be found in Ref.~\cite{Nag10}.
To encourage new experiments seeking hypernuclei, it is essential to make a 
detailed theoretical investigation of the possible existence of bound states, despite
some uncertainty in contemporary interaction models~\cite{Hiy08}.
To advance in the knowledge of the details of the $N\Xi$ interaction, 
high-resolution spectroscopy of $\Xi$ hypernuclei using $^{12}$C targets in $(K^-,K^+)$ 
reactions has been awaited~\cite{Nak10,Nat10} and it is now planned at J-PARC~\cite{Nag15}.
Identification of hypernuclei in
coming experiments at J-PARC will contribute significantly to
understand nuclear structure and baryon-baryon interactions in 
the strangeness $-2$ sector.

When a two-baryon interaction is attractive, if the system is merged with nuclear matter
and the Pauli principle does not impose severe restrictions, the attraction may be reinforced.
Simple examples of the effect of a third or a fourth baryon in two-baryon
systems could be given. The deuteron, $(I)J^P=(0)1^+$, is bound by $2.225$ MeV, while the triton,
$(I)J^P=(1/2)1/2^+$, is bound by $8.480$ MeV, and the $\alpha$ particle, $(I)J^P=(0)0^+$,
is bound by $28.295$ MeV. The binding per nucleon $B/A$ increases from $1:3:7$.
A similar argument could be employed for strangeness $-1$ systems. Whereas there is 
no evidence for dibaryon states, the hypertriton $^3_\Lambda$H, $(I)J^P=(0)1/2^+$, is bound with a separation
energy of $130 \pm 50$ keV, and the $^4_\Lambda$H, $(I)J^P=(0)0^+$, is bound
with a separation energy of $2.04 \pm 0.04$ MeV. This cooperative effect of the attraction in 
the two-body subsystems when merged in few-baryon states was also made evident
in the prediction of a $\Sigma NN$ quasibound state in the $(I)J^P = (1)1/2^+$ 
channel very near threshold~\cite{Gar07}.
Such $\Sigma NN$ quasibound state has been recently suggested
in $^3\rm{He}(K^-,\pi^\mp)$ reactions at 600 MeV/c~\cite{Har14}.
One should also bear in mind how delicate is the few-body problem in the regime
of weak binding, as demonstrated in Ref.~\cite{Nem03} for the $^4_{\Lambda\Lambda}$H
system.

It is the purpose of this paper to highlight a particular set of quantum 
numbers in the three-baryon strangeness $-2$ system that brings together
all the expected characteristics as to be bound in nature, this is the 
$(I)J^P=(3/2)1/2^+$ state. This set of quantum numbers could be achieved
by means of a $\Xi NN$ state, being decoupled from the
lowest $N\Lambda\Lambda$ state, and it would therefore be stable.
It will get two different contributions from the two-body subsystems: 
the $(i,j)=(1,0)$ $NN$ state with a spectator $\Xi$,
that will benefit from maximum coupling in isospin space preserving
the attraction of the $NN$ subsystem; and the $(i,j)=(1,0)$ and $(1,1)$ $N\Xi$ states with a spectator
$N$, that also benefit from maximum coupling in isospin space preserving
the expected attractive character of the $N\Xi$ interaction in isospin 1 partial
waves~\cite{Yam01,Hiy01,Rij13,Nag14,Car12}. Besides, this state will also gain 
additional binding due to the electromagnetic 
interaction what makes it worthwhile to look for. 

The first evidence of a deeply bound state of $\Xi^- - ^{14}$N has been recently reported~\cite{Nak15},
indicating that $\Xi-$nucleus interactions are attractive. Together with other indications of
certain emulsion data, these data suggest that the average $N\Xi$ interaction should be 
attractive~\cite{Hiy01,Rij13,Nag14}. In particular, the ESC08c Nijmegen potential for baryon-baryon 
channels with total strangeness $-2$ predicted an important 
attraction in the $i=1$ $N\Xi$ interaction, with a bound state 
of 8.3 MeV in the $(i,j)=(1,1)$ $N\Xi$ channel~\cite{Rij13}.
The recent update of the ESC08c potential to take into account the new experimental information
of Ref.~\cite{Nak15} concludes the existence of a bound state in the  
$(i,j)=(1,1)$ $N\Xi$ channel with a binding energy of 1.56 MeV~\cite{Nag14}.
The attractive character of the $i=1$ $N\Xi$ interaction has also been noticed in
the quark-model analysis of Ref.~\cite{Car12}, with a bound state of 4.8 MeV in 
the $(i,j)=(1,0)$ $N\Xi$ channel and an almost bound state in the $(i,j)=(1,1)$ $N\Xi$ channel.
In this work we will study the strangeness $-2$ three-baryon $(I)J^P=(3/2)1/2^+$ state
using existing $N\Xi$ interactions with attractive isospin 1 channels.
Thus, for the present analysis we will follow the line of the Nijmegen $N\Xi$ interaction
of Refs.~\cite{Rij13,Nag14} and also the constituent quark cluster model (CQCM) analysis of Ref.~\cite{Car12}.

In order to uncover the main features of this system we will perform first a Faddeev calculation
where the $NN$ subsystem is in the $(i,j)=(1,0)$ channel and the $N\Xi$ subsystem is either in the
$(i,j)=(1,1)$ or $(i,j)=(1,0)$ channel. We use for the $NN$ interaction the Reid soft-core potential~\cite{Rei68},
and for the $N\Xi$ interaction in any of the two channels the potential, 
\begin{equation}
V(r)=-A\, e^{-\alpha r}/r \, + \, B \, e^{-\beta r}/r \, ,
\end{equation} 
with $A=332$ MeV fm, $B=1500$ MeV fm, $\alpha=1.5$ fm$^{-1}$,  and $\beta=4.0$ fm$^{-1}$. With
this interaction the $N\Xi$ subsystem is almost bound and its phase shift changes sign at about
the same energy as the $NN$ subsystem.

Since two of the particles of this system are identical fermions, the corresponding Faddeev equations
are~\cite{Gar07}:
\begin{equation}
T=a \,t_N^{N\Xi} \, G_0 \, T \, + \, 2\, b \, t_N^{N\Xi} \, G_0 \, t_\Xi^{NN} \, G_0 \, T,
\label{eq1}
\end{equation}
where the subscript (superscript) in the two-body amplitudes $t$ denotes the spectator 
(interacting pair), and the constants $a$ and $b$ contain the spin-isospin recoupling coefficients 
and the phase arising from the reduction for identical--particles~\cite{Gar07}.

For the $\Xi NN$ three-baryon system with $J=3/2$, only the $(i,j)=(1,1)$ $N\Xi$ channel contributes 
and one finds $a=-1$, $b=0$. Thus, due to the negative sign of $a$, the $N\Xi$ interaction is 
effectively repulsive and, therefore, no bound state is possible for $J=3/2$ in spite of 
the attraction of the $N\Xi$ subsystem. The minus sign in $a$ is a consequence of the 
identity of the two nucleons since the first term of the r.h.s. of Eq.~(\ref{eq1})
proceeds through $\Xi$ exchange and it corresponds to a diagram where the initial and final states 
differ only in that the two identical fermions have been interchanged which brings the minus sign. 
This effect has been pointed out before~\cite{Gar87}.

In the case of the $\Xi NN$ three-baryon system with
$J=1/2$ one finds: $a=1/2$, $b=3/4$ for the $(i,j)=(1,1)$ $N\Xi$
channel and $a=-1/2$, $b=1/4$ for the $(1,0)$ channel. Thus, in the first case 
both terms in the r.h.s. of Eq.~(\ref{eq1}) are attractive while in the second case 
the first term is effectively repulsive and the second term is attractive, but a factor
of three smaller than that of the previous case so that effectively the $(i,j)=(1,0)$ $N\Xi$ 
channel is weakly attractive and the $(i,j)=(1,1)$ $N\Xi$ channel is the dominant one.

If we solve Eq.~(\ref{eq1}) for the $\Xi NN$ state $(I)J^P=(3/2)1/2^+$ using 
as input the $NN$ Reid soft-core potential~\cite{Rei68} and
only the $(i,j)=(1,0)$ $N\Xi$ channel, no bound state is obtained. On the 
other hand, using instead only the $(i,j)=(1,1)$ $N\Xi$ channel, we obtain a 
binding energy of 269 keV, what confirms its dominant character. If we 
now let the $(i,j)=(1,1)$ $N\Xi$ channel to become bound by increasing the 
parameter $A$ of the interaction then, as the binding energy of the two-body 
subsystem increases, the binding energy of the three-body system increases as 
well. Thus, for example, with $A=482$ MeV fm, the $(i,j)=(1,1)$ $N\Xi$ 
subsystem is bound by 8.3 MeV, similar to the ESC08c Nijmegen model~\cite{Rij13}, 
and the three-body system has a binding energy of 6.35 MeV. Similarly, 
with $A=399$ MeV fm, the $(i,j)=(1,1)$ $N\Xi$ 
subsystem is bound by 1.56 MeV, in agreement with the 
recent ESC08c Nijmegen model update~\cite{Nag14}, 
and the three-body system has a binding energy of 2.50 MeV.

We have finally performed a full-fledged Faddeev calculation~\cite{Gar13} using
the constituent quark cluster model analysis of Ref.~\cite{Car12}. 
Including only the $(i,j)=(1,1)$ $N\Xi$ channel we get a binding energy of
84 keV, while including both $N\Xi$ channels one gets a binding energy of 429 keV.
Notice that the binding energies obtained from this model are much smaller 
than those obtained from the Nijmegen-inspired model, since here the dominant channel,
the $(i,j)=(1,1)$ $N\Xi$ subsystem, is almost bound while there it has a binding energy
of 8.3 MeV~\cite{Rij13} or 1.56 MeV~\cite{Nag14}.

Let us note that current $\Xi$ hypernuclei studies~\cite{Yam10,Hiy08,Yam01} 
have been also performed by means of $N\Xi$ interactions derived from the Nijmegen interaction
models and thus our study complements such previous works for the simplest
system that could be studied exactly, the $\Xi NN$ system. One can also find in the literature 
models for the baryon-baryon interaction in the 
strangeness $-2$ sector based in EFT calculations~\cite{Pol07}, that
also show $i=1$ $N\Xi$ attraction, although one cannot conclude the strength of the
interaction due to the huge effective ranges reported.

To summarize, we have shown that using either simple phenomenological potentials or
a full-fledged Faddeev calculation with realistic $N\Xi$ interactions, derived either 
from the latest Nijmegen models or from a constituent quark cluster model,
there may exist a $\Xi$ hypernucleus with baryon number three
and quantum numbers $(I)J^P=(3/2)1/2^+$. We have highlighted the 
particular interest of the $I=3/2$ channels, because they are decoupled from the
$N\Lambda\Lambda$ state, and the $J^P=1/2^+$ state where the Pauli principle works
favorably. Besides, this state would benefit
from additional binding coming from the Coulomb potential, that in the case of the 
Nijmegen inspired models would account for a few MeV.
The equivalent $J^P=3/2^+$ state, with maximum coupling in spin and isospin space, 
could not be bound in spite of the attraction of both two-body subsystems, 
due to the phase appearing from the reduction for identical-particles in the Faddeev 
equations, that make the $N \Xi$ interaction 
effectively repulsive. This result is a consequence of the expected $N\Xi$ attraction 
in isospin 1 channels, as it occurs in the ESC08c Nijmegen 
potential~\cite{Rij13,Nag14} and the CQCM of Ref.~\cite{Car12}.
One should emphasize the importance that the $i=1$ $N\Xi$ attractions 
are strong, because in the present experimental situation the most promising production of $\Xi^-$
hypernuclei is by $(K^-, K^+)$reactions. Then, any produced $\Xi^-$ systems have to be 
in neutron--excess, because of $\Delta i_z = 1$
transfers on available nuclear targets. For such 
systems, the $i=1$ $N\Xi$ attraction works favorably.

Let us finally comment on the possible detection of this state. Out of the
four isospin components, 
\begin{eqnarray}
\left|3/2,+3/2\right> &=& pp\Xi^0 \, , \nonumber \\ 
\left|3/2,+1/2\right> &=& \frac{1}{\sqrt{3}}\left[(pn+np)\Xi^0 + pp\Xi^-\right] \, , \nonumber \\
\left|3/2,-1/2\right> &=& \frac{1}{\sqrt{3}}\left[(pn+np)\Xi^- + nn\Xi^0\right] \, , \nonumber \\
\left|3/2,-3/2\right> &=& nn\Xi^- \, , \nonumber
\end{eqnarray}
the $i_z=-1/2$ component would also benefit
from electromagnetic attraction between opposite charge particles
without penalizing electromagnetic repulsion between any of the pairs,
reenforcing the possible existence of this tribaryon beyond the attractive nuclear
contribution. In brief, this new type of element, if it does exist, would provide a great 
opportunity for extending our knowledge to some unreached part in our matter world.

\acknowledgments 
This work has been partially funded by COFAA-IPN (M\'exico), 
by Ministerio de Educaci\'on y Ciencia and EU FEDER under 
Contract No. FPA2013-47443-C2-2-P and by the
Spanish Consolider-Ingenio 2010 Program CPAN (CSD2007-00042).

\end{document}